\documentclass[a4paper,twoside,10pt]{letter}
\usepackage{graphicx,saj,multicol,subeqnarray}
\usepackage{url}

\def\papertype{Original Scientific Paper}

\def\udc{...}
\begin{document}
\baselineskip=3.1truemm
\columnsep=.5truecm
\newenvironment{lefteqnarray}{\arraycolsep=0pt\begin{eqnarray}}
{\end{eqnarray}\protect\aftergroup\ignorespaces}
\newenvironment{lefteqnarray*}{\arraycolsep=0pt\begin{eqnarray*}}
{\end{eqnarray*}\protect\aftergroup\ignorespaces}
\newenvironment{leftsubeqnarray}{\arraycolsep=0pt\begin{subeqnarray}}
{\end{subeqnarray}\protect\aftergroup\ignorespaces}
%

% Running titles

\markboth{\eightrm BEHAVIOUR OF ELECTRON CONTENT IN D-REGION DURING SOLAR X-RAY FLARES} {\eightrm M. TODOROVI\'{C} DRAKUL ET AL.}

{\ }

\publ

\type

{\ }

% Title

\title{BEHAVIOUR OF ELECTRON CONTENT IN  THE IONOSPHERIC D-REGION DURING SOLAR X-RAY FLARES}

% Authors

\authors{
        M.~Todorovi\'{c}~Drakul$^{1}$,
        V.~M.~\v{C}ade\v{z}$^{2}$,
        J.~Baj\v{c}eti\'{c}$^{3}$,
        L.~\v{C}~Popovi\'{c}$^{2}$
        D.~Blagojevi\'{c}$^{1}$
        and A.~Nina$^{4}$}

\vskip3mm

% Address

\address{$^1$Department of Geodesy and Geoinformatics, Faculty of Civil Engineering, University of Belgrade, Bulevar kralja Aleksandra 73, 11000 Belgrade, Serbia}

%\Email{mtodorovic}{grf.bg.ac.rs, bdragan@grf.bg.ac.rs}

\address{$^2$Astronomical Observatory, Volgina 7, 11060 Belgrade, Serbia}

%\Email{vcadez}{aob.rs, lpopovic@aob.rs}

\address{$^3$Department of Telecommunications and Information Science, University of Defence, Military Academy, Generala Pavla Juri\v{s}i\'{c}a \v{S}turma 33, 11000 Belgrade, Serbia}

%\Email{bajce05}{gmail.com}

\address{$^4$Institute of Physics,
University of Belgrade\break Pregrevica 118, 11080 Belgrade,
Serbia}

\Email{sandrast}{ipb.ac.rs}
% Received and Accepted dates

\dates{--}{--}

% Abstract

\summary{One of the most important parameters in ionospheric
plasma research also having a wide practical application in wireless
satellite telecommunications is the total electron content (TEC) representing
the columnal electron number density. The F-region with high electron density provides the
biggest contribution to TEC while the relatively weakly ionized
plasma of the D-region (60 km - 90 km above Earth's surface)
is often considered as a negligible cause of satellite signal disturbances.
However, sudden intensive ionization processes like those induced by
solar X-ray flares can cause relative increases of electron density
that are significantly larger in the D-region than in regions at
higher altitudes. Therefore, one cannot  exclude a
priori the D-region from investigations of
ionospheric influences on propagation of electromagnetic signals emitted by satellites. We discuss here this problem which has not been
sufficiently treated in literature  so far. The obtained results are
based on data collected from the D-region monitoring by very low
frequency radio waves and on vertical TEC calculations from the Global
Navigation Satellite System (GNSS) signal analyses, and they show
noticeable variations in the D-region's electron content (TEC$_{\mathrm{D}}$)
during activity of a solar X-ray flare (it rises by a factor of
136 in the considered case) when TEC$_{\mathrm{D}}$ contribution to
TEC can reach several percent and which cannot be neglected
in practical applications like global positioning procedures by
satellites.}

% Keywords (see keywords.pdf file)


\begin{multicols}{2}
{

% Sections

\section{1. INTRODUCTION}

The terrestrial ionosphere is used for detection and investigations
of different effects produced by numerous phenomena in
outer space and different Earth layers. The research in this field
includes scientific analyses of various phenomena in the
ionosphere as e.g. lightnings (Inan et al. 2010), tectonic motions
(Astafyeva and Afraimovich 2006, Dautermann et al. 2009), the
sunrise and sunset (Afraimovich et al. 2009, Nina and \v{C}ade\v{z} 2013),
solar eclipses (Singh et al. 2011, Maurya et al. 2014, Verhulst et
al. 2016), $\gamma$ ray burst (Inan et al. 2007, Nina et al. 2015a), solar
activity (Kolarski et al. 2011, Nina et al. 2011, 2012a,b, Schmitter
2013, \v{S}uli\'{c} and Sre\'{c}kovi\'{c} 2014, Satori et al. 2016), nuclear explosions
(Madden and Thompson 1965, Yang et al. 2012, Zhang and Tang 2015) as well as ionospheric
harmonic and quasi-harmonic hydrodynamic motions (Jilani et al.
2013, Nina and \v{C}ade\v{z} 2013). However, the ionosphere has also a very
important role in practical applications, primarily in
telecommunications (Baj\v{c}eti´c et al. 2015, Alizadeh et al.
2015).

The total electron content (TEC) is one of the most important
parameters that describes the ionospheric state and structure and
provides an overall reflection of ionization processes. It
is the column electron number density defined as the number of free
electrons in a column of unit cross section extending from the
ground to the top of the ionosphere. In addition to scientific
research (see, for example, Afraimovich et al. 2009) TEC is used in
calculations needed for practical applications in telecommunication
(for example in systems which use transionospheric radio waves,
Hofmann-Wellenhof et al. 2001) and in global navigation satellite
systems (GNSS, Jakowski et al. 2005). Keeping in mind that radio
communications, positioning, navigation and timing by GNSS
signals play a critical role in telecommunications, geodesy and land
surveying, emergency response, precision agriculture, all forms of
transportation (space stations, aviation, maritime, rail, road and
mass transit) the studies of relevant signal propagations and
consequently TEC is of great importance at the present time.

TEC measurements are made mostly using the GNSS data because of a
good global coverage of the GNSS observation network (Shim 2009).
However, TEC can be calculated from electron densities obtained by
different methods depending on altitude. For the low-altitude
ionospheric monitoring they are based on radio propagations, and rocket and radar
measurements (see Grubor et al. 2005, Strelnikova and Rapp
2010, Nina et al. 2011, Chau et al. 2014, Kolarski and Grubor 2014) while the
upper ionospheric monitoring is based on ionosonde and satellite measurements
(see Stankov et al. 2011).

The most important contribution to TEC during standard
unperturbed conditions is the ionospheric F-region while the contribution from the lowest ionospheric altitudes can be neglected. However,
variations in perturbation's intensity can cause significant changes of
physical properties of the ionosphere affecting plasma parameters, Schuman resonance, propagation of radio signals in a wide range of frequency bandwidths (ULF, SLF, ELF, VLF, LF, HF), etc. which is treated in literature (Kulak et al.  2003, Satori et al. 2005, 2016, Eccles et al. 2005, Williams and Satori 2007, Nina et al. 2015.)
 As investigations indicate,
intensity of perturbations can be strongly dependent on location which, as a result,
may change the contribution of a particular layer of the
ionosphere to the total TEC. A phenomenon which may require a
modification of the calculation procedures applicable in the case of
unperturbed conditions is a solar X-ray flare.
Namely, the  relative increase of radiation following such an event is of more
intensive influence in the lower than in the upper ionosphere in daytime conditions. This is
concluded in numerous investigations showing that not too intensive
X-ray flares produced practically no changes to TEC (described primarily by the
F-region plasma) while, at the same time, the electron density increase in the
D-region altitude can be raised by more than one order of magnitude
(Nina et al. 2012a,b, Singh et al.
2014). Also, during a very intensive increase of X radiation the height
of the electron density maximum can be lowered to the E-region
(Xiong et al. 2011).

The main aim of this study is to examine whether the total electron content in the D-region (TEC$_{\mathrm{D}}$) i.e. the column electron number density of the
D-region may have a contribution to TEC that can no more be ignored in calculations if the ionosphere gets perturbed by solar X-ray flares. We are focused on analyses of variations in the D-region electron
content and its contribution to
the total TEC during a solar X-ray flare. Here, we point out that the presented study is related to the daytime ionosphere when the influence of a solar X-ray flare is observable. During nighttime conditions there are no visible signal reactions to flare events.

Our research has two parts. First, we study the time
evolution of TEC$_{\mathrm{D}}$ and its contribution to TEC for a particular
event. Keeping in mind that X-ray flares induce similar ionospheric
changes in space and time, and that the goal of this study is to
present a procedure for relevant calculations, we consider one
typical representative flare of class C8.8 which appeared on May
5th, 2010 and was detected by the GOES-14 satellite. To monitor the
D-region and to compute TEC$_{\mathrm{D}}$ we use a technique based on the VLF/LF signal propagation relevant to this region. We analyse the DHO VLF signal emitted in
Germany and received in Serbia while TEC data are taken from the
website http://www.bath.ac.uk/elec-eng/invert/iono/rti.html. The
second part of this paper contains a statistical analysis of
 the dependence of TEC$_{\mathrm{D}}$ on this class of solar X-ray flares e.i. on their maximum
intensity.
%The main goal of our work is prediction of TEC$_D$ and its contribution in TEC for X-ray flares of some maximum intensity detected by satellite.

\section{2. EXPERIMENTAL SETUP AND OBSERVATIONAL DATA}
\label{exp}

Calculations are here based on data collected by three
experimental setups for monitoring the solar X-ray radiation and
ionosphere.

First, the considered time period is chosen using data for photon flux recorded by the National Oceanic and Atmospheric
Administration (NOAA) satellite GOES-14 (\url{http://satdat.ngdc.noaa.gov/sem/goes/data/new_full/2010/05/goes14/csv/g14_xrs_2s_20100505_20100505.csv}). Here, our detailed research of changes in the contribution of the D-region to vertical TEC was directed to ionospheric perturbations induced by the solar X-ray
flare between 11:37 UT and 11:58 UT, May 5, 2010 (Nina and \v{C}ade\v{z} 2014, Nina et al. 2015b) recorded as increase of solar radiation within the wavelengths range 0.1 nm - 0.8 nm which is shown in Fig. 1, in the
upper panel.

Second, data for the D-region observation were obtained using the 23.4 kHz VLF
signal emitted by the DHO transmitter located in Rhauderfehn
(Germany) and recorded by the receiver at the Institute of Physics in Belgrade
(Serbia). The
ionospheric perturbations were detected as amplitude $ A_{\mathrm
{rec}}$ and phase $ P_{\mathrm {rec}}$ variations of the considered VLF
signal recorded by the AWESOME (Atmospheric Weather Electromagnetic
System for Observation Modelling and Education) VLF receiver (Cohen
et al. 2010) (Fig. 1, two middle panels). The DHO transmitter was chosen because it provides the strongest signal received in Belgrade which propagates along
a relatively short path (the entire path of around 1300 km is over Europe i.e. it is not transoceanic)
allowing us to assume a practically uniformly stratified space within the observational time period which is important for applied modeling.

Third, we calculate the mean TEC related to the considered area
using data available on the website
\url{http://www.bath.ac.uk/elec-eng/invert/iono/rti.html}. These
data are determined from the GPS (Global Positioning System e.i. US
GNSS) measurement and MIDAS (Multi-Instrument Data Analysis System)
tomographic algorithm (Mitchell and Spencer 2003). The working range
covers an area stretching from the northwest encompassing
Rhauderfehn (Germany) to the southeast and including the area of
Belgrade (Serbia).

\section{3. D-REGION MODELLING}
\label{modelling}

Modeling the signal propagation requires knowledge of the electron density dependence on space and time.
In this paper, the electron density during an X-ray flare is determined by a procedure given in Grubor et al. (2008) and used in many other papers (see for example \v{Z}igman et al. 2007, Kolarski et al. 2011, Nina et al. 2011, 2012a,b). This procedure is based on matching the observed VLF signal data, amplitude $\Delta A_{\mathrm{rec}}$ and phase $\Delta P_{\mathrm{rec}}$ changes, with corresponding results of simulations $\Delta A_{\mathrm{sim}}$ and $\Delta P_{\mathrm{sim}}$
  of the VLF signal propagation using the Long-Wave Propagation Capability (LWPC) numerical model developed by the Naval Ocean Systems Center (NOSC), San Diego, USA (Ferguson 1998). The registered amplitude $\Delta A_{\mathrm{rec}}$ and phase $\Delta P_{\mathrm{rec}}$ changes are determined with respect to corresponding values in unperturbed conditions. In our case they are 25.6 dB and 16.2${^\mathrm{{\circ}}}$ respectively as seen in Fig. 1, two middle panels.
As to the simulated values of amplitude and phase, they are obtained using Wait's model of the horizontally stratified ionosphere (Wait and Spies 1964) which is incorporated in the LWPC numerical model of the VLF/LF signal propagation, and is characterized by two independent parameters: the "sharpness" $\beta$ and the signal reflection height $H^\prime$. These parameters are the inputs for the LWPC program which then computes $A_{\mathrm{sim}}(\beta,H')$ and $P_{\mathrm{sim}}(\beta,H')$ i.e. their corresponding changes $\Delta A_{\mathrm{sim}}(\beta,H')$ and $\Delta P_{\mathrm{sim}}(\beta,H')$ as output values. Finally, the parameters $\beta$ and $H^\prime$ are determined from the condition of best fitting the experimental data with their numerical counterparts:

\[
\displaystyle \Delta A_{\mathrm{sim}}(\beta,H')\approx \Delta A_{\mathrm{rec}}(t),
\]
 \begin{equation}
\label{eq::uporedjivanje}
\displaystyle  \Delta P_{\mathrm{sim}}(\beta,H')\approx \Delta P_{\mathrm{rec}}(t).
\end{equation}
According to the above relation Eq. (1), the obtained pairs of Wait's parameters are time dependent $\beta=\beta(t)$ and $H^\prime=H^\prime(t)$ as presented in Fig. 1 (the bottom panel) for the considered case. The electron density now follows from Wait's equation as:

 \begin{equation}
\label{eq::Ne}
\displaystyle N_{\mathrm e}(h,t)
= 1.43\cdot10^{13}{\rm e}^{-\beta(t)H^\prime(t)}{\rm e}^{(\beta(t)-\beta_0)h},
\end{equation}
where $N_{\mathrm e}$ in m$^{-3}$, $H^\prime$(t) and $h$ are in km, $\beta$ is
in km$^{-1}$ and $\beta_0=0.15$ km$^{-1}$. The general expression of
TEC$_{\mathrm{D}}$:

\begin{equation}
\label{eq::DEC1} {\mathrm{TEC_D}}(t) = \int_{h_{\mathrm b}}^{h_{\mathrm t}}N_{\mathrm e}(h,t)dh
\end{equation}
has the following dimensional form (in m$^{-2}$) if Eq.(2) is taken into account:

\[
{\mathrm{TEC_D}}(t) = 1.43\cdot 10^{16}{\rm e}^{-\beta(t)H^\prime(t)}
\int_{h_b}^{h_t}{\rm e}^{(\beta(t)-\beta_0)h}dh
\]
\begin{equation}
\label{eq::DEC2}
=1000\frac{N_{\mathrm e}(h_{\mathrm t},t)-N_{\mathrm e}(h_{\mathrm b},t)}{\beta(t)-\beta_0},
\end{equation}
where $h_{\mathrm b}$ = 60 km and $h_{\mathrm t}$ = 90 km are the bottom and upper D-region boundary, respectively.

\centerline{\includegraphics[bb=0 0 629
815,width=\columnwidth, keepaspectratio]{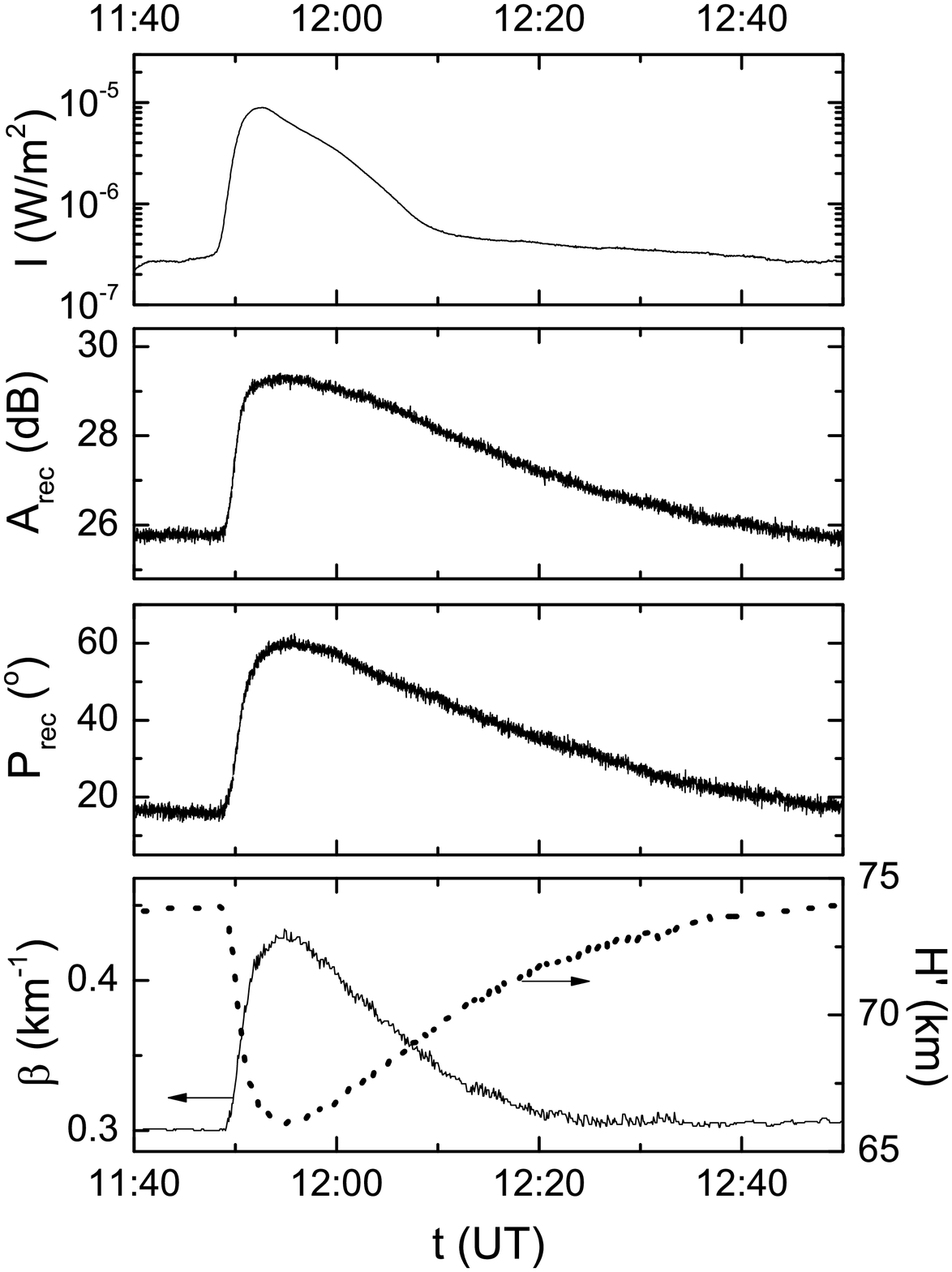}}

\figurecaption{1.}{Time dependencies of the X radiation intensity
registered by the GOES-14 satellite in the wavelength domain between
0.1 and 0.8 nm (top panel), signal amplitude and phase (middle
panel) registered by the AWESOME receiver located in Belgrade
(Serbia), and the calculated Wait's parameters (bottom panel) during
the presence of the solar X-flare on May 5, 2010 (Nina and
\v{C}ade\v{z} 2014).}

Relative contribution of TEC$_{\mathrm{D}}$($t$) to TEC($t$) can now be defined as:

\begin{equation}
\label{eq::rDEC} r_{\mathrm D}(t) =\frac{{\mathrm{TEC_D}}(t)}{{\mathrm {TEC}(t)}}.
\end{equation}

Finally, to analyse fractional contributions of particular sublayers
of the D-region in TEC$_{\mathrm D}$ changes we calculate $\Delta$TEC$_{\mathrm {Di}}$
for the $i$-th sublayer and obtain analogously to Eq. (4):

\begin{equation}
\label{eq::DECi} \Delta {\mathrm {TEC_{Di}}}(t) =1000
\frac{N_{\mathrm e}(h_{\mathrm {i+1}},t)-N_{\mathrm e}(h_{\mathrm i},t)}{\beta(t)-\beta_0}
\end{equation}
where:
\begin{equation}
\label{eq::hi}
h_{\mathrm i} = h_{\mathrm b} + (i-1)\delta h,\quad \mbox{i=1,...,i$_{\mathrm {max}}$}
\end{equation}
is the height of the lower boundary of the i-th sublayer, $i_{\mathrm {max}}=15$ is the total number of considered sublayers taken with equal thickness $\delta h=2$ km so that $h_{\mathrm t}=h_{\mathrm b}+i_{\mathrm {max}}\delta h$.

\section{4. RESULTS AND DISCUSSION}

As we said in the Introduction, the two main issues of this study are
the analyses of the temporal evolution of TEC$_{\mathrm{D}}$ characteristics during
a solar X-ray flare affecting the Earth's atmosphere, and
the maximum X-ray intensity influence on TEC$_{\mathrm{D}}$.

\subsection{4.1. Time evolution of TEC$_{\mathrm{D}}$ characteristics}

To study the temporal evolution of TEC$_{\mathrm{D}}$ we used data extracted
from the DHO VLF signal registered by  the AWESOME receiver system
in Belgrade at the time of the solar X-ray flare influence on the
ionosphere that occurred on May 5, 2010, and applied the procedures
explained in Section 3 to calculate relevant quantities. This event
represents a case of a clear response of the D-region to the
considered perturbation alone as no other possible source of
intensive disturbance was noticeable at the time (see signal
reactions in Fig. 1). For this reason this event was also treated in
our earlier papers (Nina and \v{C}ade\v{z} 2014, Nina et al. 2015b).
In addition, this flare is not too intensive (class C8.8) and, as it
will be shown in the analysis presented in the second part of this
investigation, it can produce smaller considered changes in TEC$_{\mathrm{D}}$
than events followed by larger intensities. So this analysis is also
related to examination of the need for inclusion of the low
ionosphere in modeling  small TEC changes. This is important for
practical application because X-ray flares of class C (X-ray emission in 0.1 nm - 0.8 nm with maximum intensity of 10$^{-6}$ W/m$^2$ - 10$^{-5}$ W/m$^2$) are more
frequent than those of classes M and X (X-ray emission in 0.1 nm - 0.8 nm with maximum intensity of 10$^{-5}$ W/m$^2$ - 10$^{-4}$ W/m$^2$ and above 10$^{-4}$ W/m$^2$, respectively). This is visible during the entire solar cycle period. For example, the GOES satellite recorded 28 and 0 events of C and M class solar X-ray flares, respectively, in 2009 (the solar cycle minimum), while the corresponding numbers were 1797 and 207 in 2014 (the solar cycle maximum).

\vskip-2cm

\centerline{\includegraphics[bb=0 0 629 815,width=\columnwidth,
keepaspectratio]{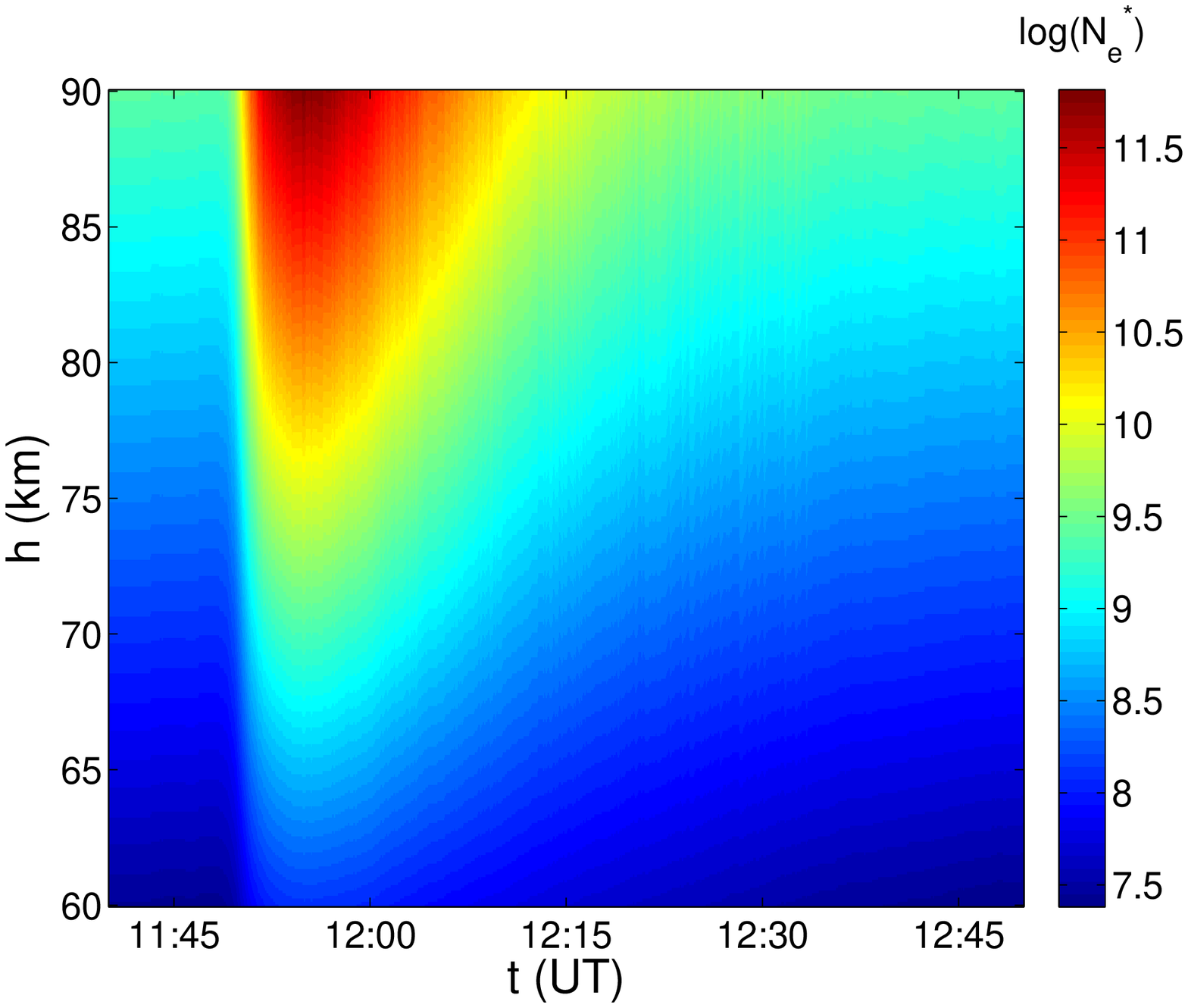}}
\vskip-2cm

\figurecaption{2.}{Surface plot of log$(N_e^*(t,h))$ as a function of time $t$ and altitude $h$ during the considered solar X-ray flares where $N_e^*=N_e/N_e^0$, where $N_e^0=1$ m$^{-3}$.}

As we can see from Eq. (3), determination of TEC$_{\mathrm{D}}$ requires
knowledge of the electron density distribution with altitude. So, we
first applied Wait's equation Eq. (2) with a set of Wait's
parameters $\beta$ and $H^\prime$ (Fig. 1, the bottom panel)
calculated by the established methodology described in Section 3.
The altitude and temporal dependence of the resulting modeled
electron density $N_{\mathrm e}(h,t)$ is shown in Fig. 2. Here it can be seen
that the most pronounced electron density variation with height
occurs at the time of the electron density maximum around 11:54 UT
when it rises by three orders of magnitude, and that the maximum
increase relative to the initial value is obtained at 90 km (by more
than factor 250). These properties of the electron concentration
yield  TEC$_{\mathrm{D}}$: its time evolution and its internal vertical
structure arising from individual horizontal layers as illustrated
in Figs. 3 and 4.

The time evolution of TEC$_{\mathrm{D}}$ obtained from Eq. (4) and presented in
Fig. 3 (upper panel) shows that it increases from 0.0017 TECU to
0.2302 TECU (1 TECU = $10^{16}$ m$^{-2}$) rising by a factor 136. At the same time, the
corresponding data for TEC  obtained from the web
(\url{http://www.bath.ac.uk/elec-eng/invert/iono/rti.html}, see
Section 2) show its rise from 5.74245 at 11:45 UT (before the flare)
to 6.04857 at  12:00 UT which is several minutes after the maximum
perturbation of the D-region due to the rather coarse  time
resolution of 15 min used in TEC data collection. In spite of this,
it is now obvious that TEC increases   percentually significantly
less than TEC$_{\mathrm{D}}$ (see Table 1) which is clearly visible in Fig. 2,
the bottom panel, showing the contribution of TEC$_{\mathrm{D}}$ to TEC given
by $r_{\mathrm{D}}=\mathrm{TEC}_{\mathrm{D}}/\mathrm{TEC}$ which rises from 0.03\% to a maximum of not less than 1.5\% (such an estimate results from the available time sampling of TEC data which is much coarser than in the case of TEC$_{\mathrm{D}}$) within the same
time interval.

\vskip.5cm \noindent
\parbox{\columnwidth}{
{\bf Table 1.} TEC$_{\mathrm{D}}$, TEC and $r_{\mathrm{D}}$ during the D-region
disturbances. \vskip.25cm \centerline{\begin{tabular}{cccc} \hline
t (UT) & TEC$_{\mathrm D}$ (TECU)& TEC (TECU)& $r_{\mathrm{D}}$\\
\hline
11:45 & 0.00165 & 5.74245 & 0.02868\\
12:00 & 0.09191 & 6.04857 & 1.51959\\
12:15 & 0.00612 & 5.51302 & 0.11107\\
12:30 & 0.00221 & 6.98893 & 0.03157\\
12:45 & 0.0018 & 6.21602 & 0.029\\
\hline
\end{tabular}}}
 \vskip.5cm

\vskip-5cm

\centerline{\includegraphics[bb=0 0 629
815,width=\columnwidth, keepaspectratio]{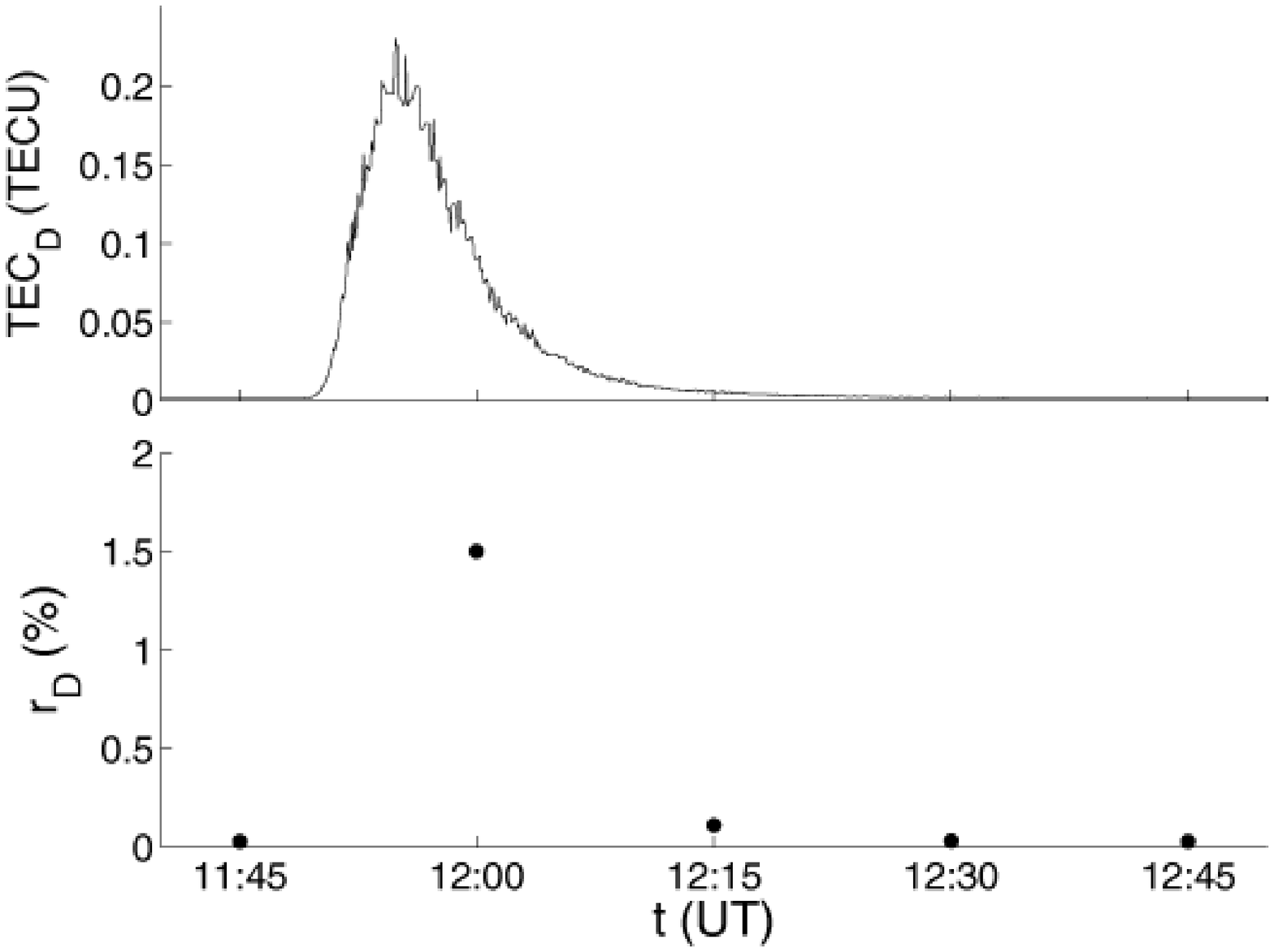}}

%\vskip-2cm

\figurecaption{3.}{Time evolutions of TEC$_{\mathrm{D}}$ (upper panel) and its
contribution to TEC (bottom panel) during the considered solar X-ray
flare.}

To analyse contributions to TEC$_{\mathrm{D}}$ coming from different heights
within the D-region, e.g. to investigate the influence of the
considered  solar flare X-radiation on different parts of the D-region, we divide it into horizontal layers of equal
thickness $\Delta h=2$ km and calculate the related time dependence
for the partial $\Delta$TEC$_{\mathrm{Di}}(t)$ for  60 km $< h <$ 90 km. Fig. 4 shows that changes are more intensive at the top of the D-region which is in agreement with  Fig. 2 for the  electron density dependence.

However, the perturbation variations with altitude are seen more
clearly by looking at relative increases of $\Delta$TEC$_{\mathrm{Di}}(t)$
with respect to some stationary value for unperturbed conditions
$\Delta$TEC$_{\mathrm{Di0}}$:

%\vskip-4cm
\begin{equation}
\label{eq::Dtec}
r^{\Delta}_i (t)=\frac{ \Delta \mathrm{TEC}_{\mathrm {Di}}(t)-\Delta \mathrm{TEC}_{\mathrm {Di0}}}{\Delta \mathrm{TEC}_{\mathrm {Di0}}}
\end{equation}
where $\Delta$TEC$_{\mathrm{Di0}}$ is obtained from Eq. (4) using
characteristic Wait's parameters for the unperturbed ionosphere,
according to Grubor et al. 2008. Fig. 5 shows that $r(t)$  varies
substantially with height $h$ only during the period of the most
intense D-region plasma perturbations when they reach two orders of
magnitude.

\vskip-5cm

\centerline{\includegraphics[bb=0 0 629
815,width=\columnwidth, keepaspectratio]{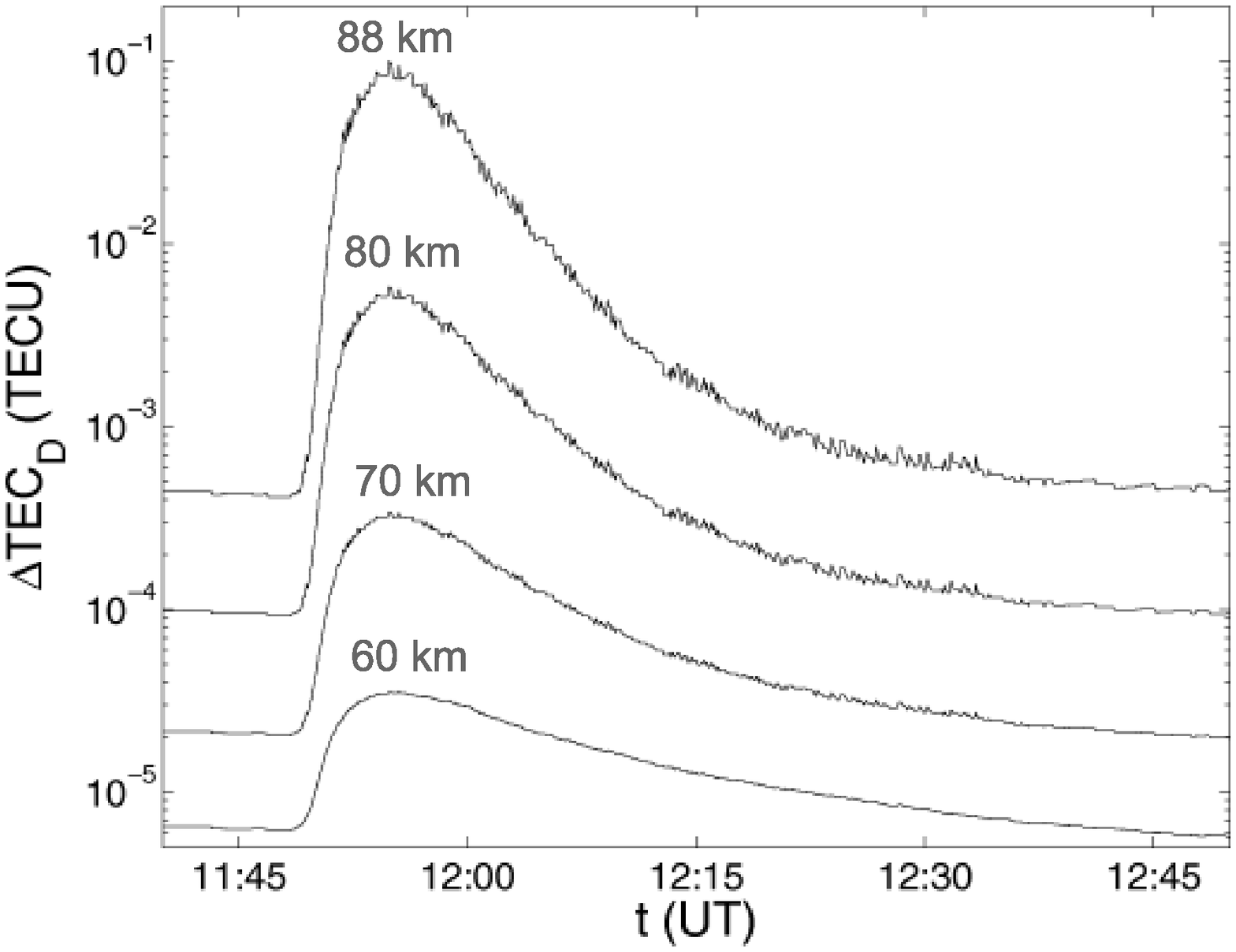}}

%\vskip-2cm

\figurecaption{4.}{Time evolution of $\Delta$TEC$_D$ in layers (of thickness of 2 km) located at altitudes 60 km, 70 km, 80 km, and 88 km during the considered solar X-ray flare.}
\vskip-3cm

\centerline{\includegraphics[bb=0 0 629 815,width=\columnwidth,
keepaspectratio]{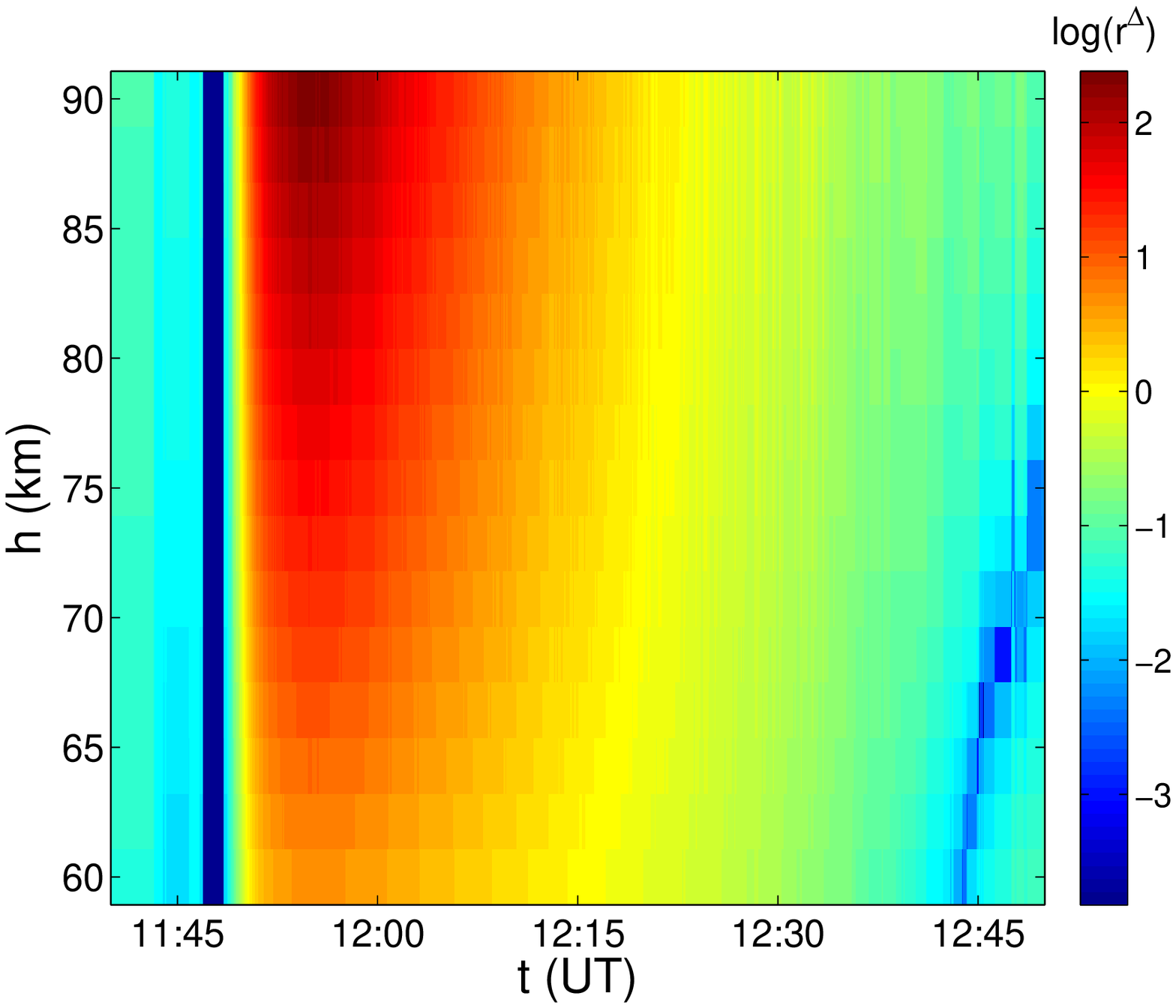}}

\vskip-3cm

\figurecaption{5.}{Surface plot
of relative changes of $\Delta TEC_{{D}}(t,h)$ with respect to
relevant values related to unperturbed conditions as a function of
time $t$ and altitude $h$ given by Eq. (8) during the considered solar X-ray flare.}

\vskip 5cm

\subsection{4.2. TEC$_{\mathrm{D}}$ characteristics at the X-ray intensity maximum}

In investigations of the TEC$_{\mathrm{D}}$ dependence on maximal intensities of
flares radiations $I_{\mathrm {max}}$ we used statistical analyses for
different flare events from Thompson et al. 2005 and Grubor et al.
2008. The calculation of TEC$_{\mathrm {D}}$ at the maximum X-ray intensity is
based on Wait's parameters $\beta$ and $H'$ for the analysed events. We first fit
these parameters by expressions:
\begin{equation}
\label{eq::rDEC}
\beta(I_{\mathrm {max}})=C_1+C_2{\mathrm {log}}(I_{\mathrm {max}}^*)+C_3{\mathrm {log}}(I_{\mathrm {max}}^*)^2
\end{equation}
and
\begin{equation}
\label{eq::rDEC}
H^\prime(I_{\mathrm {max}})=D_1+D_2{\mathrm {log}}(I_{\mathrm {max}}^*)
\end{equation}
(as shown in Fig. 6) where $I_{\mathrm {max}}^*=I_{\mathrm {max}}/I_0$, $I_0=1$ W/m$^{-2}$ and
 coefficients $C_1$, $C_2$, $C_3$, $D_1$, $D_2$ are given in Table 2.
These fits are then used in calculations of $N_{\mathrm e}$ and TEC$_{\mathrm{D}}$ following the procedure from Section 3. The obtained dependencies of TEC$_{\mathrm{D}}$ on
$I_\mathrm{{max}}$ for flare classes C and M are given in Fig. 7  showing a
significant increase in TEC$_{\mathrm{D}}$ with $I_{\mathrm {max}}$. Keeping in mined that these values are
significantly larger than those for the unperturbed ionosphere we
used them as changes of TEC$_{\mathrm{D}}$ induced by the solar X-ray flare at
its intensity maximum. For the C class
flares TEC$_{\mathrm{D}}$ ranges between about 0.01 TECU and 0.1 TECU at the
considered time, and it reaches several TECU for the M class flares
which affects the propagation of electromagnetic waves and
becomes very important for practical use  of GNSS signals in different
measurements. In this figure it can also be noticed that  data from
Thompson et al. (2005) yield larger TEC$_{\mathrm{D}}$ than data from Grubor et
al. (2008). Such a difference is expected because the former data are related
to the lower latitude D-region where the local ionizing solar radiation is stronger
resulting in a larger local TEC$_{\mathrm{D}}$ in comparison with the middle latitude D-region
where the local radiation is weaker.

%
%\vskip.5cm \noindent
%\parbox{\columnwidth}{
%{\bf Table 1.} DEC, TEC and $r_D$ during the D-region disturbances. \vskip.25cm
%\centerline{\begin{tabular}{cccccc}
%\hline
%Data source & $C_1$ & $C_2$& $C_3$& $D_1$& $D_2$\\
%\hline
%Grubor et al. 2008 & 0.3872 & -0.0841 & -0.0154 & 48.0166 & -3.7381\\
%Thomson et al. 2005 & 0.4916 & -0.0385 & -0.0095 & -4.8976 & 42.1210\\
%\hline
%\end{tabular}}}
% \vskip.5cm

\vskip.5cm \noindent
\parbox{\columnwidth}{
{\bf Table 2.} Fitted coefficients used in Eqs. (9) and (10). \vskip.25cm
\centerline{\begin{tabular}{ccc}
\hline
Data source & Grubor et al. (2008) & Thomson et al. (2005)\\
\hline
$C_1$ &   0.3872 &  0.4916\\
$C_2$&-0.0841 &-0.0385 \\
$C_3$&-0.0154 &-0.0095\\
$D_1$&48.02 &42.12 \\
$D_2$&-3.7381&-4.8976\\
\hline
\end{tabular}}}
 \vskip -2cm

\centerline{\includegraphics[bb=0 0 629
815,width=\columnwidth, keepaspectratio]{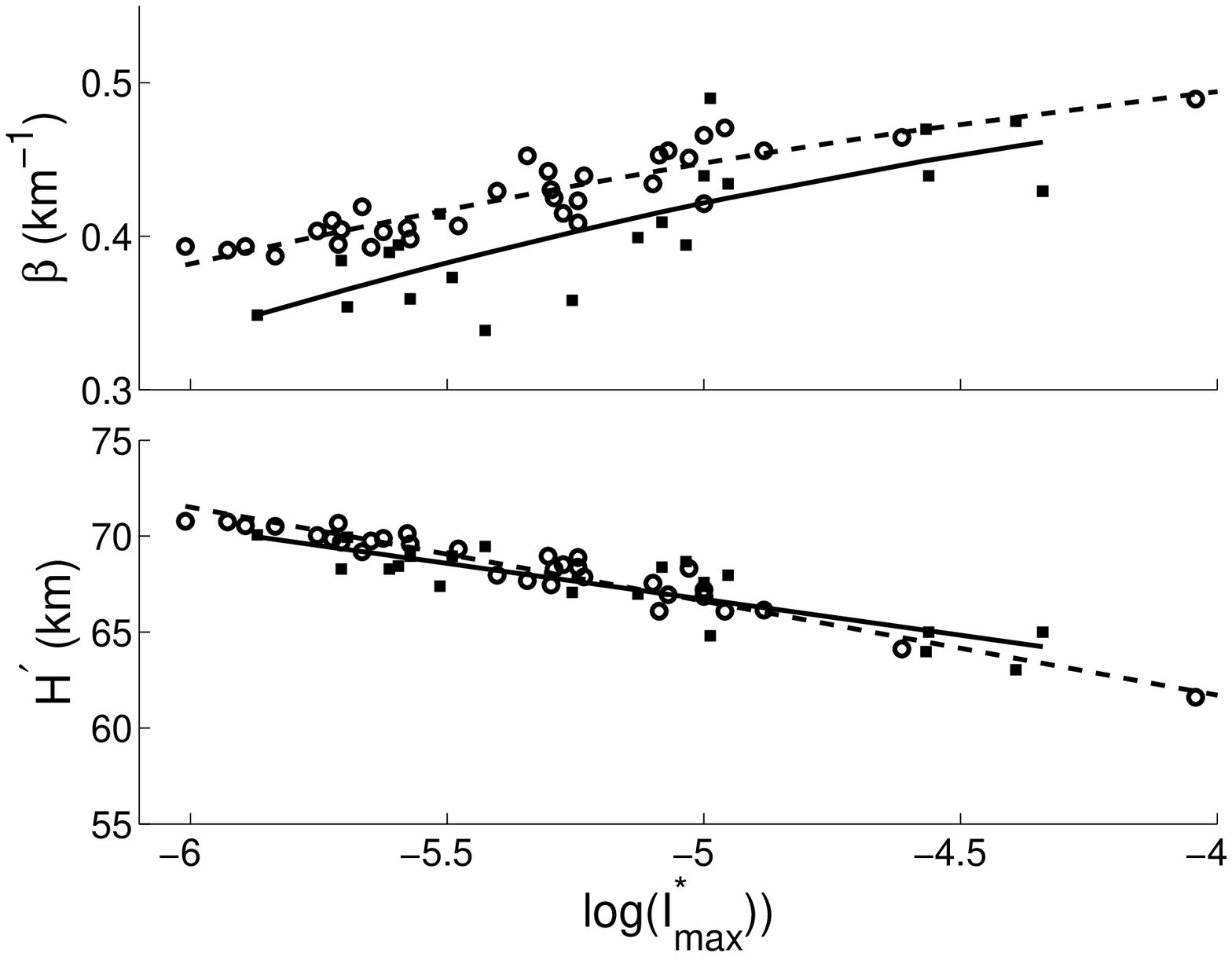}}

 \vskip -3cm
\figurecaption{6.}{Wait's parameters at the times of maxima in solar X-ray flares given in Grubor et al. (2008) (filled squares) and Thomson et al. (2005) (open circles) as a function of $I_{max}^*=I_{max}/I_0$, where $I_0=1$ W/m$^2$. Relevant fitted curves expressed by Eqs. (9) and (10) are shown as solid and dashed lines for the first and second initial data sets, respectively.}

%To estimate the contribution of DEC changes $\Delta DEC$ in TEC changes $\Delta TEC$ we taken two expressions Eq. (9) and Eq. (10) that describe TEC dependencies versus $I_{max}$ given in Afraimovich et al. 2002 and Liu et al. 2006 , respectively:
%\begin{equation}
%\label{eq::rDEC}
%\Delta TEC_1=649.4 \cdot I^{0.7}
%\end{equation}
%and
%\begin{equation}
%\label{eq::rDEC}
%\Delta TEC_2=0.102+7490 \cdot I
%\end{equation}
%and plot them in Fig. 7, middle panel. Finally, we calculate ratio of both DEC1 and DEC2 and both TEC1 and TEC2 (Fig. 7, bottom panel). Here it is important to point out that all ionospheric changes depend on the considered location. In this paper we give only estimations of the considered parameters and calculat all 4 combinations and detailed analyses related to the geographical position is in focus of our upcoming research. The obtained results shown increase in contribution of DEC in TEC with $I_{max}$

\vskip-5cm

\centerline{\includegraphics[bb=0 0 629
815,width=\columnwidth, keepaspectratio]{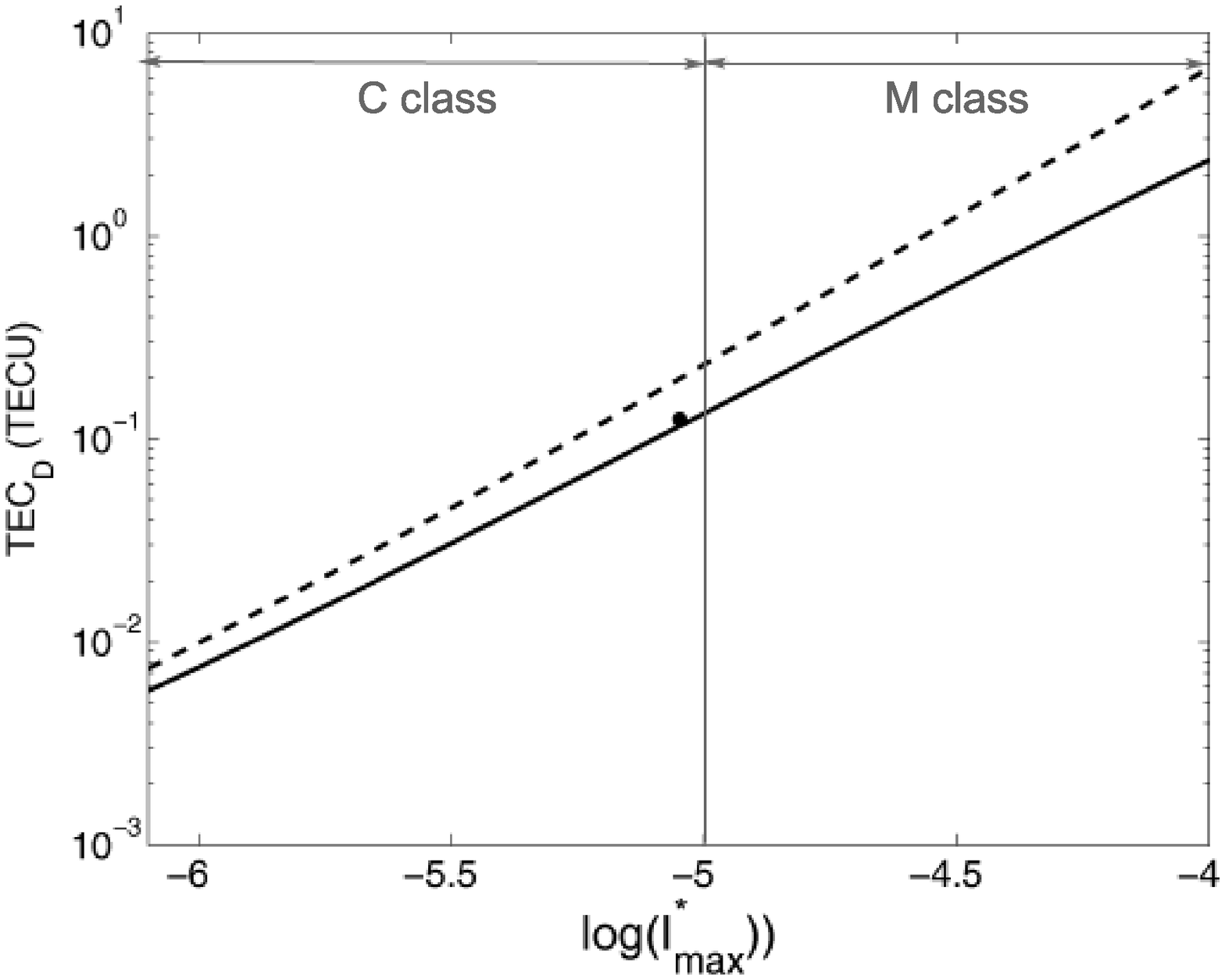}}

%\vskip-2cm

\figurecaption{7.}{Dependencies of TEC$_D$ in the times of X-ray
intensity maxima of C and M classes flare events calculated from
data given in Grubor et al. (2008) (solid line) and Thomson et al.
(2005) (dash line) as a function of $I_{max}^*=I_{max}/I_0$, where $I_0=1$ W/m$^2$.}

\section{5. CONCLUSIONS}

In this paper we investigate the D-region electron content changes
induced by solar X-ray flares. The presented analysis consists of
two parts:
\item{$\bullet$} First, we analyse time evolutions of TEC$_{\mathrm{D}}$ and its
contribution to TEC as well as contributions of different layers of
the horizontal D-region parts during a chosen flare disturbation.
The method of the study is based on data obtained from the DHO VLF
signal emitted in Germany and collected by the VLF receiver located
in Serbia. We also used data for the TEC given on the website
\url{http://www.bath.ac.uk/elec-eng/invert/iono/rti.html}.
\item{$\bullet$} Second,
we analyse the influence of flare intensity on changes in TEC$_{\mathrm{D}}$
and its contribution in TEC variations  applying our  technique to a
number of different events using their data from literature.

To summarize,  the obtained results show:

\item{(i)} TEC$_{\mathrm{D}}$ can increase by more than two orders of magnitude (in the considered case of intensive flare it increased from 0.0017 TECU to 0.2302 TECU which is equivalent to an increase by a factor of 136).
\item{(ii)} The relative increase of $\Delta$TEC$_{\mathrm{D}}$ with respect to its initial values is the most pronounced during the perturbation maximum in the D-region plasma when that reaches two orders of magnitude in the considered case.
\item{(iii)} Variations of the TEC$_{\mathrm{D}}$ contribution in TEC are significantly larger during the maximum perturbation with respect to the unperturbed condition. In the considered case, it increases from 0.03\% to more then 1.5\% at several minutes after the perturbation maximum
\item{(iv)} A significant increase of TEC$_{\mathrm{D}}$ with I$_{\mathrm {max}}$ is seen at the time of the maximum in X-radiation intensity, $I=I_{\mathrm {max}}$ for different flares. The values obtained using LWPC numerical model for propagation VLF/LF radio waves based on Wait's model of ionosphere indicate that TEC$_{\mathrm{D}}$ takes values from 0.01 TECU to 0.1 TECU for the class C flares at considered time, while they reach several TECU in the case of the class M flares.

From these conclusions we can see that ionization variations in the D-region induced by solar X-ray flares become important in modeling the GNSS signal propagations and its practical applications in measurements during flare activity. The influence of flares significantly increases with $I_{\mathrm {max}}$ and  it can not be neglected especially in accurate measurements like those in geodesy.

Finally, we want to point out that the present study opens numerous questions related to the D-region influence on effective ionospheric
effects on GNSS signal propagations during intensive perturbations  induced by solar X-ray flares.

\acknowledgements{The authors would like to thank the Ministry of Education, Science and Technological Development of the Republic of Serbia for the support of this work within the projects III-44002, 176001, 176002, 176004 and TR36020. The authors are also grateful to an anonymous
referee for very useful suggestions and comments.
The data for this paper collected by GOES-14 satellite
is available at NOAA's National Centers for Environmental information (\url{http://satdat.ngdc.noaa.gov/sem/goes/data/new_full/2010/05/goes14/csv/g14_xrs_2s_20100505_20100505.csv}). The mean TEC related to the considered area
are obtained from the website
\url{http://www.bath.ac.uk/elec-eng/invert/iono/rti.html}.
 Requests for the VLF
 data used for analysis can be directed to the corresponding author.
}

\references

{Afraimovich}, E.~L., {Edemskiy}, I.~K., {Leonovich}, A.~S., {Leonovich}, L.~A.,
  {Voeykov}, S.~V. and {Yasyukevich}, Y.~V.: 2009, \journal{Geophys. Res. Lett.}, \vol{36}(15), 106.

  Alizadeh, M. M., Schuh, H., Schmidt, M.: 2015, \journal{Radio Sci.},
\vol{50}(6), 539.

{Astafyeva}, E.~I. and {Afraimovich}, E.~L.: 2006, \journal{Earth Planets Space}, \vol{58}, 1025.

{Baj\v{c}eti\'{c}}, J., {Nina}, A., \v{C}ade\v{z}, V.~M. and Todorovi\'{c},
  B.~M.: 2015,
\journal{Therm. Sci.}, \vol{19}, Suppl. 2, S229.

Chau, J.~L., R\"{o}ttger, J. and Rapp, M.: 2014,
\journal{J. Atmos. Sol.-Terr. Phy.}, \vol{118}, 113.

{Cohen}, M., {Inan}, U. and {Paschal}, E.~W., P.: 2010,
\journal{IEEE T. Geosci. Remote.}, \vol{48}, 3.

{Dautermann}, T.,{Calais}, E.,{Lognonn{\'e}}, P. and {Mattioli} G.~S.: 2009, \journal{Geophys. J. Int.}, \vol{179}, 1537.

Eccles, J. V., Hunsucker, R. D., Rice, D. and Sojka J. J.: 2005,
\journal{Space Weather}, \vol{3}(1), S01002.

{Ferguson}, J.~A.: 1998,
Computer Programs for Assessment of Long-Wavelength Radio
  Communications, Version 2.0, Space and Naval Warfare Systems Center, San Diego.

{Grubor}, D.~P., {{\v S}uli{\'c}}, D.~M. and {{\v Z}igman}, V.: 2005,
\journal{Serb. Astron. J.}, \vol{171}, 29.

{Grubor}, D.~P., {{\v S}uli{\'c}}, D.~M. and {{\v Z}igman}, V.: 2008,
\journal{Ann. Geophys.}, \vol{26}, 1731.

Hofmann-Wellenhof, B., Lichtenegger, H. and Collins, J.: 2001, Global Positioning System: Theory and Practice, Springer-Verlag, New York.

%Inan, U.~S., Shafer, D.~C., Yip, W.~Y. and Orville, R.~E.: 1988,
%\journal{J. Geophys. Res.-Space},
%  \vol{93}, 11455.

{Inan}, U.~S., {Lehtinen}, N.~G., {Moore}, R.~C., {Hurley}, K.,{Boggs}, S.,
  {Smith}, D.~M. and {Fishman}, G.~J.: 2007, \journal{Geophys. Res. Lett.}, \vol{34}, L08103,

{Inan}, U.~S., {Cummer}, S.~A. and {Marshall}, R.~A.: 2010,
\journal{J. Geophys. Res.-Space}, \vol{115}, A00E36.

Jakowski, N., Stankov, S. M., Klaehn, D.: 2005, \journal{Ann.
Geophys.}, \vol{23}(9), 3071.

{Jilani}, K., {Mirza}, A.~M. and {Khan}, T.~A.: 2013,
\journal{Astrophys. Space Sci.}, \vol{344}, 135.

{Kolarski}, A., {Grubor}, D. and {{\v S}uli{\'c}}, D.: 2011,
\journal{Balt. Astron.}, \vol{20}, 591.

Kolarski, A. and Grubor, D.: 2014, \journal{Adv. Space Res.},
\vol{53}(11), 1595.

Kulak, A., Zieba, S., Micek, S. and Nieckarz, Z.: 2003, \journal{J. Geophys. Res}, \vol{108}(A7), 1270.

Madden, T and Thompson, W.: 1965, \journal{Rev. Geophys.}, \vol{3}(2), 211.

Maurya, A.~K., Phanikumar, D.~V., Singh, R., Kumar, S., Veenadhari, B., Kwak,
  Y.-S., Kumar, A., Singh, A.~K. and Niranjan~Kumar, K.: 2014,
\journal{J. Geophys. Res.-Space}, \vol{119}(10), 8512.

 Mitchell, C. N. and Spencer, P. S. J.: 2003,  \journal{Ann. Geophys.}, \vol{46}(4), 687.

{Nina}, A., {{\v C}ade{\v z}}, V., {Sre{\'c}kovi{\'c}}, V.~A. and {{\v
  S}uli{\'c}}, D.: 2011,
\journal{Balt. Astron.}, \vol{20}, 609.

{Nina}, A., {{\v C}ade{\v z}}, V., {Sre{\'c}kovi{\'c}}, V. and {{\v
  S}uli{\'c}}, D. 2012a,
\journal{Nucl. Instrum. Methods in Phys. Res. B},
  \vol{279}, 110.

{Nina}, A., {{\v C}ade{\v z}}, V., {{\v S}uli{\'c}}, D., {Sre{\'c}kovi{\'c}},
  V. and {{\v Z}igman}, V.: 2012b,
\journal{Nucl. Instrum. Methods in Phys. Res. B},
  \vol{279}, 106.

Nina, A. and \v{C}ade\v{z}, V.~M.: 2013,
\journal{Geophys. Res. Lett.}, \vol{40}(18), 4803.

{Nina}, A. and {{\v C}ade{\v z}}: 2014,
\journal{Adv. Space Res.}, \vol{54}(7), 1276.

{Nina}, A., Simi{\' c}, S., {Sre{\'c}kovi{\'c}} V. and {Popovi{\'c}}
L.\v{C}.: 2015a, \journal{Geophys. Res. Lett.}, \vol{42}(19), 8250.

{Nina}, A., \v{C}ade\v{z}, V.~M. and {Baj\v{c}eti\'{c}}, J.: 2015b, \journal{Serb. Astron. J.}, \vol{191}, 51.

Satori, G., Williams, E. and Mushtak, V.: 2005, \journal{J. Atmos. Sol.-Terr. Phy.}, \vol{67}(6), 553.

Satori G., Williams E., Price C., Boldi R., Koloskov A., Yampolski Y., Guha A., Barta V.: 2016, \journal{Surv. Geophys.}, \vol{37}, 757.

Schmitter, E.~D.: 2013,
\journal{Ann. Geophys.}, \vol{31}(4), 765.

Shim, JA~S.: 2009, Analysis of Total Electron Content (TEC) Variations in the Low- and Middle-Latitude Ionosphere, PhD Dissertation, Utah State University, Logan, Utah.

{Singh}, R., {Veenadhari}, B., {Maurya}, A.~K., {Cohen}, M.~B., {Kumar}, S.,
  {Selvakumaran}, R., {Pant}, P., {Singh}, A.~K. and {Inan}, U.~S.: 2011,
\journal{J. Geophys. Res.-Space}, \vol{116}, 10301.

Singh, A.~K., Singh, A., Singh, R. and Singh, R.: 2014,
\journal{Astrophys. Space Sci.}, \vol{350}(1), 1.

Stankov, S.M., Stegen, K., Muhtarov, P. and Warnant, R.: 2011,
\journal{Adv. Space Res.}, \vol{47}(7), 1172.

{Strelnikova}, I. and {Rapp}, M.: 2010,
\journal{Adv. Space Res.}, \vol{45}(2), 247.

{\v{S}uli\'{c}}, D. and {Sre\'{c}kovi\'{c}}, V.~A.: 2014,
\journal{Serb. Astron. J.}, \vol{188}, 45.

 Thomson, N. R., Rodger, J. C., and Clilverd, A. M.: 2005, \journal{J. Geophys.
Res.}, \vol{110}, A06306.

Verhulst, T.G.W., Sapundjiev, D., Stankov, S. M.: 2016,
\journal{Adv. Space Res.}, in press.

{Wait}, J.~R. and {Spies}, K.~P.: 1964,
Characteristics of the Earth-ionosphere waveguide for VLF radio
  waves, NBS Technical Note 300, National
Bureau of Standards, Boulder, CO.

Williams, E. R. and Satori, G.: 2007, \journal{Radio Sci.}, \vol{42}, RS2S11.

Xiong, B., Wan, W., Liu, L., Withers, P., Zhao, B., Ning, B.,
Wei, Y., Le, H., Ren, Z., Chen, Y. He, M. and Liu J.: 2011, \journal{J. Geophys.
Res.}, \vol{116}, A11317.

Yang, Y. M., Garrison, J. L. and Lee S. C.: 2012,  \journal{Geophys. Res. Lett.}, \vol{468}(39), L02103.

Zhang, X. and Tang, L.: 2015,
\journal{Ann. Geophys.}, \vol{33}(1), 137.

\endreferences
}

\end{multicols}

\vfill\eject

{\ }

% Serbian abstract

% Title

\naslov{OSOBINE SADR{\ZZ}AJA ELEKTRONA U JONOSFERSKOJ D-OBLASTI
TOKOM SUN{\CH}EVE ERUPCIJE U \textit{X}-PODRU{\CH}JU}

% Authors

\authors{
        M.~Todorovi\'{c}~Drakul$^{1}$,
        V.~M.~\v{C}ade\v{z}$^{2}$,
        J.~Baj\v{c}eti\'{c}$^{3}$,
        L.~\v{C}~Popovi\'{c}$^{2}$
        D.~Blagojevi\'{c}$^{1}$,
        and A.~Nina$^{4*}$}

\vskip3mm

% Address

\address{$^1$Department of Geodesy and Geoinformatics, Faculty of Civil Engineering, University of Belgrade, Bulevar kralja Aleksandra 73, 11000 Belgrade, Serbia}

%\Email{mtodorovic}{grf.bg.ac.rs, bdragan@grf.bg.ac.rs}

\address{$^2$Astronomical Observatory, Volgina 7, 11060 Belgrade, Serbia}

%\Email{vcadez}{aob.rs, lpopovic@aob.rs}

\address{$^3$Department of Telecommunications and Information Science, University of Defence, Military Academy, Generala Pavla Juri\v{s}i\'{c}a \v{S}turma 33, 11000 Belgrade, Serbia}

%\Email{bajce05}{gmail.com}

\address{$^4$Institute of Physics,
University of Belgrade\break Pregrevica 118, 11080 Belgrade,
Serbia}

\Email{sandrast}{ipb.ac.rs}

%\authors{B. Arbutina$^{1}$, D. Uro{\v s}evi{\' c}$^{1}$ and Z. Kne{\v z}evi{\' c}$^2$}
%\vskip3mm
%
%% Address
%
%\address{$^1$Department of Astronomy, Faculty of Mathematics,
%University of Belgrade\break Studentski trg 16, 11000 Belgrade,
%Serbia}
%
%\Email{arbo}{math.rs, dejanu@math.rs}
%
%
%\address{$^2$Astronomical Observatory, Volgina 7, 11060 Belgrade 38, Serbia}
%
%\Email{zoran}{aob.rs}

\vskip.7cm

% UDC

\centerline{UDK \udc}

% Papertype

\centerline{\rit Originalni nuqni rad}
%\centerline{\rit Uredjivaqki prilog}

\vskip.7cm

\begin{multicols}{2}
{

% Abstract

{\rrm Jedan od najzna{\ch}ajnijih parametara plazme koji takodje ima
i {\sh}iroku prakti{\ch}nu primenu u be{\zz}i{\ch}nim satelitskim
telekomunikacijama je tzv. ukupni sar{\zz}aj elektrona (eng.} total electron content - TEC) {\rrm koji predstavlja ukupan broj elektrona u zadanoj cilindri{\ch}noj zapremini jedini{\ch}nog popre{\ch}nog preseka. Najve{\cc}i doprinos TES-u daje} F-{\rrm oblast zbog
velike elektronske gustine dok relativno slabo jonizovana plazma u}
D-{\rrm oblasti (60 km - 90 km iznad povr{\sh}ine Zem{\lj}e) se {\ch}esto smatra kao zanemarljiv uzro{\ch}nik
perturbacija elektromagnetnog signala. Medjutim, trnutni intenzivni procesi, kao oni
indukovani Sun{\ch}evim erupcijama u \textit{X}-podru{\ch}ju,  mogu uzrokovati relativni porast
elektronske gustine koji je srazmerno znatno ve{\cc}i u }D-{\rrm oblasti nego na ve{\cc}im
visinama. Zbog toga se ne mo{\zz}e unapred isklju{\ch}iti
}D-{\rrm oblast u prou{\ch}avanjima jonosferskog uticaja
na propagacije elektromagnetnih signala emitovanih sa satelita. Ovde razmatramo
taj problem koji, do sada, nije dovoljno obra{\dj}en u literaturi.
Dobijeni rezultati su bazirani na podacima prikupljenim posmatranjem
}D-{\rrm oblasti radio talasima vrlo niskih frekvencija i prora{\ch}unima
TES-a iz analiza signala u Globalnom Navigacionom Satelit}c{\rrm kom
Sistemu (GNSS). Rezultati koje smo dobili pokazuju vidljive varijacije u elektronskom sadr{\zz}aju u }D-{\rrm oblasti} (TEC$_{\mathrm{D}}$) {\rrm tokom Sun{\ch}evih erupcija u \textit{X}-podru{\ch}ju
 kada udeo } TEC$_{\mathrm{D}}$ {\rrm u TES mo{\zz}e dosti{\cc}i i
nekoliko procenata a koji se ne mogu zanemariti u prakti{\ch}nim
primenama kao {\sh}to su, recimo, procedure globalnog satelitskog pozicioniranja.} }

\end{multicols}

\end{document}